\documentclass[conference]{IEEEtran}

\usepackage{amsmath,amssymb}
\usepackage{booktabs}
\usepackage{array}
\usepackage{enumitem}
\usepackage{xcolor}
\usepackage{listings}
\usepackage{graphicx}
\usepackage[hidelinks]{hyperref}
\usepackage{algorithm}      
\usepackage{algpseudocode}  

\definecolor{codegray}{gray}{0.96}
\definecolor{codecomment}{rgb}{0.30,0.45,0.30}
\definecolor{codekey}{rgb}{0.20,0.20,0.55}
\lstdefinestyle{qdag}{
  basicstyle=\scriptsize\ttfamily,
  backgroundcolor=\color{codegray},
  commentstyle=\color{codecomment}\itshape,
  keywordstyle=\color{codekey}\bfseries,
  columns=fullflexible,
  keepspaces=true,
  breaklines=true,
  breakatwhitespace=false,
  showstringspaces=false,
  frame=single,
  framesep=4pt,
  xleftmargin=4pt,
  xrightmargin=2pt,
  aboveskip=6pt,
  belowskip=4pt,
  morecomment=[l]{\#},
}
\newcommand{\code}[1]{\texttt{#1}}
\newcommand{\yes}{\ensuremath{\checkmark}}
\newcommand{\partialmark}{\ensuremath{\sim}}
\newcommand{\no}{\textendash}
\newcommand{\qdag}{QDAG}

\begin{document}

\title{QDAG: Declarative Composition of Reusable Analytics Methodologies at LinkedIn}

\author{
\IEEEauthorblockN{Peter Ho}
\IEEEauthorblockA{\textit{LinkedIn Corporation}\\
tinyam@gmail.com}
\and
\IEEEauthorblockN{Praveen Chaganlal}
\IEEEauthorblockA{\textit{LinkedIn Corporation}\\
pchaganlal@linkedin.com}
\and
\IEEEauthorblockN{Tianle Zhang}
\IEEEauthorblockA{\textit{LinkedIn Corporation}\\
tianzhang@linkedin.com}
\and
\IEEEauthorblockN{Ming Liang}
\IEEEauthorblockA{\textit{LinkedIn Corporation}\\
miliang@linkedin.com}
}

\maketitle

\begin{abstract}
\normalfont
Production analytics products often depend on reusable
methodologies: multi-step definitions such as headcount growth,
top-skill growth, or differentially-private impression distributions.
Although these methodologies define business-critical numbers,
they are commonly implemented as imperative glue around OLAP
queries, service calls, joins, transformations, and conditional logic.
As a result, teams duplicate orchestration code, definitions drift
across products, and methodologies are difficult to test or analyze.

We present QDAG, a production system at LinkedIn that represents
an analytics methodology as a declarative directed acyclic graph of
typed steps. Nodes may execute Apache Pinot queries, downstream
service calls, in-memory SQLite joins, jq transformations,
conditionals, differentially-private aggregations, or calls to other
QDAGs. The engine evaluates graphs demand-driven, memoized,
pruned, and parallelized in the per-request analytics mid-tier.
QDAG is deployed across more than 500 hosts with over 340 \qdag{}s defined
across more than 100 production use cases, adding roughly 10 ms median orchestration
overhead and under 50 ms at the 99th percentile. Our experience
shows that making methodologies declarative improves reuse,
testability, and cross-product consistency while preserving
interactive latency.
\normalfont
\end{abstract}
\medskip



\section{Introduction}
\label{sec:intro}

Every analytics product is, at bottom, a collection of \emph{definitions}. A
talent-intelligence product must agree on what ``year-over-year headcount growth
for a company in a region'' means; a content-analytics product must agree on what
``unique impressions of a post'' means. Teams
call these definitions \emph{methodologies}. A methodology is the part of an
analytics application that carries business meaning, that auditors ask about, and
that two products must share if their numbers are to agree. Methodologies are the
computation layer beneath the metrics, dashboards, and features that data-mining
and machine-learning products consume: when two products disagree on ``headcount
growth,'' a methodology has drifted, and every analysis, ranking, and model built
on it inherits the error.

A methodology is almost never a single query. Consider three real examples drawn
from production analytics surfaces at LinkedIn:

\begin{itemize}[leftmargin=1.2em,itemsep=2pt,topsep=2pt]
  \item \textbf{Headcount growth.} Run one OLAP query for current headcount per
  country; run a second query for the prior period, \emph{constrained to the
  countries the first query returned}; join the two result sets and compute a
  growth ratio.
  \item \textbf{Top skills.} Select the top-$N$ skills with one query, then issue
  \emph{further} queries parameterized by those selected skills to compute a
  growth rate per skill -- a data dependency between queries with a transform in
  between to extract the entities.
  \item \textbf{Unique post impressions.} Issue two \emph{composition of differentially-private query results}
  aggregate queries (a per-dimension count and a total) \emph{in parallel}, drop
  non-positive rows, and divide to obtain a percentage distribution -- collapsing
  two separate downstream calls into one reusable result.
\end{itemize}

Each of these is a small dataflow program over a fast OLAP store. Today that
program is written by hand: SQL strings assembled by a bespoke query-builder
module, wrapped in imperative service code that sequences and parallelizes the
queries, post-processes JSON, and branches on conditions. The result is three
recurring problems, observed across multiple lines of business:

\begin{enumerate}[leftmargin=1.4em,itemsep=2pt,topsep=2pt]
  \item \textbf{Authoring cost.} Multi-query methodologies require real
  concurrency and orchestration code. Going from ``I have the SQL'' to ``I have a
  correct, parallel, conditionally-branching endpoint'' is the expensive step.
  \item \textbf{Drift and misalignment.} The same methodology is re-implemented in
  slightly different ways across endpoints and teams, and the copies diverge.
  Definitions that ought to be identical quietly stop matching.
  \item \textbf{Per-team frameworks.} Each line of business builds its own
  query-composition framework (one in Scala, several in Java), duplicating
  engineering effort and making cross-team alignment nearly impossible.
\end{enumerate}

\qdag{} (Query Directed Acyclic Graph)\footnote{Internally the system is 
referred to as the Enhanced Query
Processor - EQP. We use \qdag{} throughout for the model and ``the \qdag{} engine'' for
the implementation.} is our response. A methodology is declared once, as a graph
whose nodes are typed steps and whose edges are data dependencies, and is then
executed by a shared engine. The engine -- not the application author -- decides
what to run, in what order, what to run in parallel, what to skip, and what to
reuse. \qdag{} is deployed in LinkedIn's analytics mid-tier and is in production
use today.

\paragraph{What this paper is.}
This is an experience paper about a deployed system. \qdag{} combines established
techniques -- demand-driven, memoized, conditionally-pruned, parallel DAG
evaluation from \texttt{make} and modern build
systems~\cite{feldman,bsalc,bazel,shake}, and the ``define a metric once and reuse
it'' goal of the semantic-layer
movement~\cite{metricflow,malloy,lookml,cube} -- and applies them at a point in the
design space they have not previously occupied together: the per-request OLAP
mid-tier. Our contributions are:

\begin{itemize}[leftmargin=1.2em,itemsep=2pt,topsep=2pt]
  \item A declarative model (Section~\ref{sec:model}) that unifies OLAP queries,
  in-process relational joins, JSON transforms, differentially-private
  aggregation, downstream service calls, runtime conditionals, and recursive
  composition into one artifact for analytics methodologies.
  \item An engine (Section~\ref{sec:engine}) that brings the build-system
  evaluation model -- demand-driven, memoized, pruned, parallel -- into the
  \emph{per-request OLAP mid-tier} at interactive latency, deployable both
  embedded in a client and remotely in a gateway, with a mock mode that makes
  methodologies unit-testable without touching a data store.
  \item A surface-syntax design (Sections~\ref{sec:jq}--\ref{sec:sql}): a
  YAML-embedded macro layer that compiles to jq, and a dynamic SQL-template plus
  predicate DSL for safe, composable query construction.
  \item A positioning relative to federation engines, semantic layers, and
  workflow schedulers (Section~\ref{sec:related}), and deployment experience
  (Section~\ref{sec:experience}) across more than 100 production use cases.
\end{itemize}

\section{Background and Motivation}
\label{sec:motivation}

\paragraph{What counts as a methodology.}
We use \emph{methodology} in a specific sense, and the boundary matters for what
\qdag{} does and does not target. A methodology is a multi-step, parameterized
computation that produces a business-meaningful result: it has a fixed structure and
varying inputs, it composes more than one step, and its correctness is a matter of
agreement across teams rather than of local implementation. This distinguishes it
from two neighbors. A \emph{metric} is a single value -- headcount, impressions --
that a methodology may produce; metrics are the outputs, methodologies the
computations that yield them, and a single methodology often produces several
related metrics. A \emph{data product} is the surface that consumes those results
-- a dashboard, an API, a ranking feature; data products are \qdag{}'s callers, not
its subject. Concretely, the methodologies we target share four characteristics:
they issue between two and roughly a dozen steps; at least one step depends on the
result of another; they mix relational and non-relational work (an OLAP query, an
in-process join, a JSON reshape); and they run in a user request, so the whole
computation must complete in milliseconds. Single-query lookups do not need
\qdag{}, and offline pipelines are better served by a batch scheduler
(Section~\ref{sec:related}); the target is the space between.

\paragraph{The substrate.}
The analytics surfaces we target are backed by Apache Pinot~\cite{pinot}, a
distributed real-time OLAP store designed for low-latency aggregation queries over
very large fact tables. Like other real-time OLAP stores -- Druid~\cite{druid} and
ClickHouse~\cite{clickhouse}, in the interactive-analytics lineage of
Dremel~\cite{dremel} -- Pinot answers a single query quickly, but it does not
\emph{orchestrate} a methodology: it has no notion of ``run these queries, join
their results, branch on a condition, and post-process'' across multiple
round-trips. That orchestration lives above Pinot, in a mid-tier service.

\begin{figure}[t]
\centering
\includegraphics[width=0.92\columnwidth]{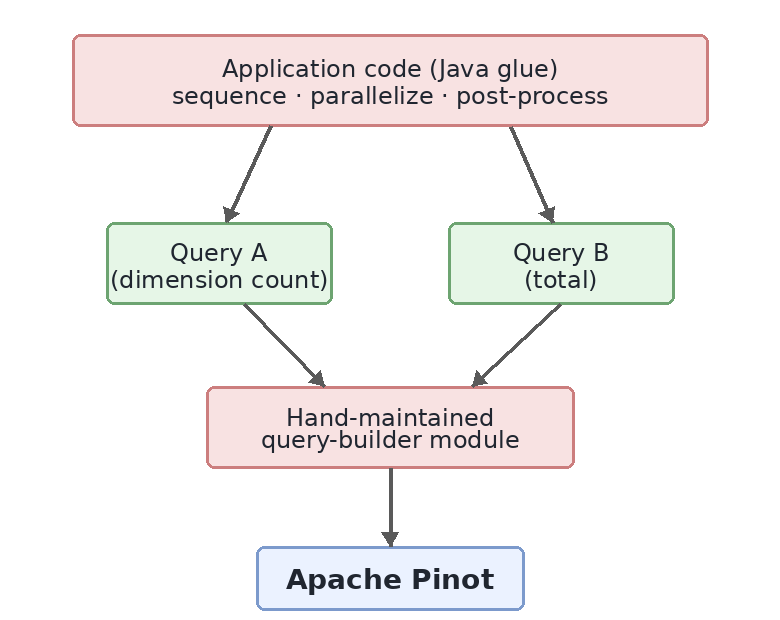}
\caption{The status quo for a two-query methodology: orchestration,
parallelization, and post-processing are hand-written application glue (red),
sitting on a per-team query-builder module. The glue is what drifts and
duplicates.}
\label{fig:before}
\end{figure}

\paragraph{The status quo.}
Figure~\ref{fig:before} sketches the conventional implementation of the
unique-post-impressions methodology. Two endpoints each issue a query through a
general-purpose, hand-maintained query-builder module -- routed through a separate
differential-privacy service for noise injection, in the manner of LinkedIn's
PriPeARL framework~\cite{pripearl,dwork-roth} -- the caller makes two network
round-trips, and the application code drops non-positive rows and computes
percentages in Java. Every non-query box is application-level glue, and every line of glue is
written, tested, and maintained by the application team. When Pinot gains a new
SQL feature (for example a \code{HAVING} clause), the query-builder module must be
extended to expose it -- maintenance that each team pays separately.

\paragraph{Why a declarative graph.}
Three properties of methodologies make a declarative graph attractive. First,
their \emph{structure is stable but their parameters vary}: the shape of
``headcount growth'' does not change per request, only the region does. Second,
they are \emph{compositional}: ``growth rate'' is reused inside many higher-level
definitions. Third, they are \emph{analyzable}: if the methodology is data rather
than code, an engine can validate it today and, in principle, estimate its
resource footprint and reason about its latency before it runs.(Section~\ref{sec:analyzable}). Imperative glue
forfeits all three. A declarative artifact recovers them -- which is precisely the
argument the semantic-layer community makes for metrics, and which we extend below
the metric, to the procedural composition of heterogeneous steps.

\section{The \qdag{} Model}
\label{sec:model}

A \qdag{} is a named, acyclic graph of \emph{nodes}. Each node is a typed
execution unit that consumes and produces a single JSON value; an edge from node
$a$ to node $b$ means $b$ depends on $a$'s result. Exactly one node is designated
the \emph{target}; execution begins there and proceeds only as far as the target's
transitive dependencies require. An implicit input value, addressable as
\code{\$QDAG\_INPUT}, is injected into every invocation, so the same graph
parameterizes over per-request inputs.

\subsection{A worked example}
\label{sec:example}

\begin{figure}[t]
\centering
\includegraphics[width=0.82\columnwidth]{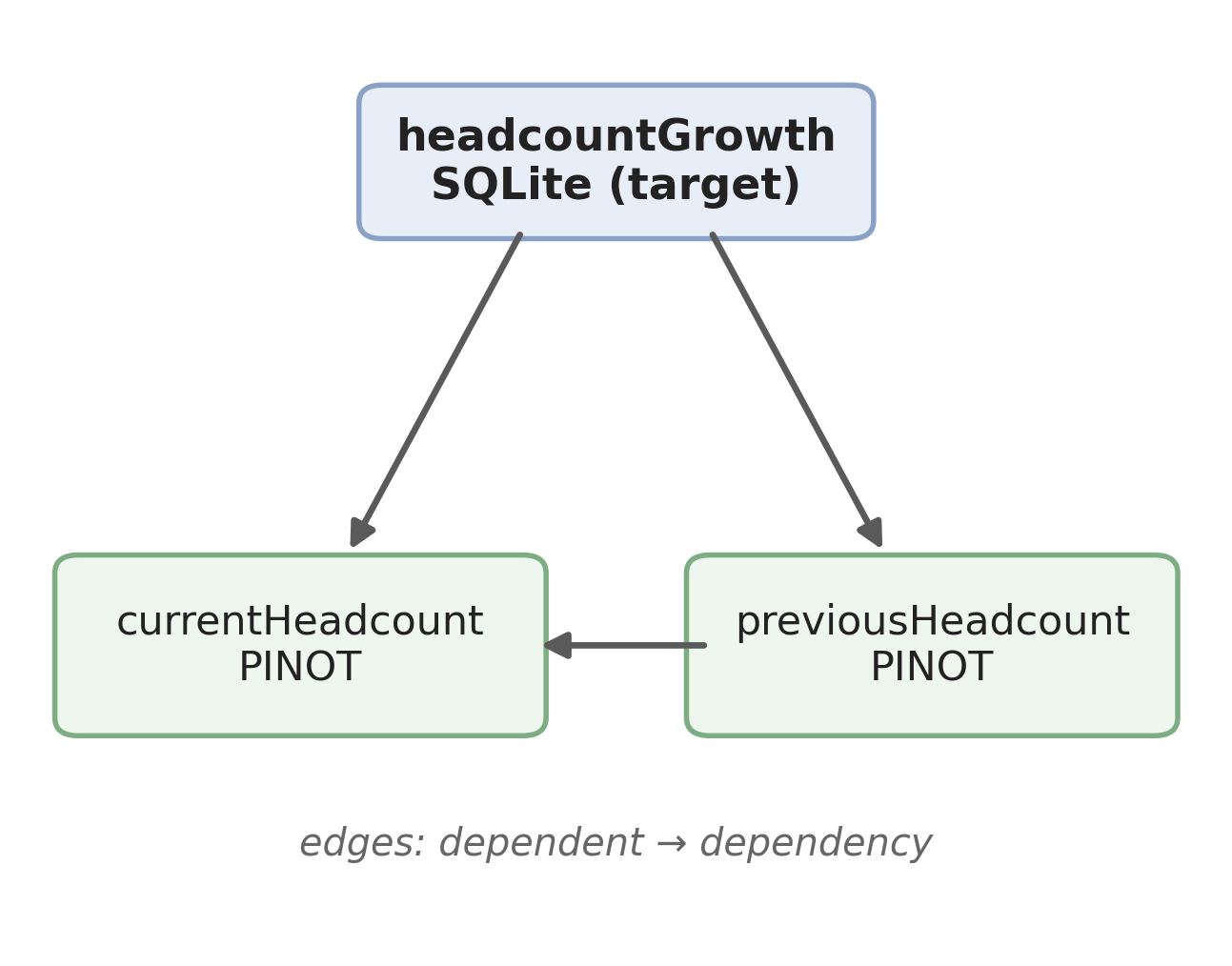}
\caption{Dependency graph of the \texttt{HeadcountGrowth} \qdag{}. Edges are drawn from a node to the nodes it depends on; that is,
dependent -> dependency. \code{currentHeadcount} is a shared dependency of both other nodes and
is evaluated once.}
\label{fig:headcount}
\end{figure}

Figure~\ref{fig:headcountyaml} shows the complete ``headcount growth''
methodology and Figure~\ref{fig:headcount} its dependency graph. It has three
nodes. \code{currentHeadcount} runs a Pinot query for the
current period. \code{previousHeadcount} runs a second Pinot query for the prior
period, but \emph{depends on} \code{currentHeadcount}: it constrains its
\code{WHERE} clause to exactly the countries the first query returned, by
splicing them into an \code{IN} list. \code{headcountGrowth}, the target, is a
SQLite node that joins the two result sets in process and computes the growth
ratio. Note that \code{currentHeadcount} is referenced by \emph{both}
\code{previousHeadcount} and \code{headcountGrowth}; the engine executes it once
and reuses the result (Section~\ref{sec:engine}).

\begin{figure}[t]
\centering
\begin{lstlisting}
name: HeadcountGrowth
nodes:
  currentHeadcount:
    type: PINOT
    d2: currentMemberTableService
    sql: |
      SELECT countryCode, count(memberId) AS headcount
      FROM currentMemberTable
      WHERE continentCode = {{$QDAG_INPUT.continentCode}}
      GROUP BY countryCode
      ORDER BY count(memberId) DESC LIMIT 3

  previousHeadcount:
    type: PINOT
    d2: previousMemberTableService
    dependencies: [currentHeadcount]
    sql: |
      SELECT countryCode, count(memberId) AS headcount
      FROM previousMemberTable
      WHERE continentCode = {{$QDAG_INPUT.continentCode}}
      AND countryCode IN
        ({{[$currentHeadcount.data[].countryCode]}})
      GROUP BY countryCode
      ORDER BY count(memberId) DESC LIMIT 3

  headcountGrowth:
    type: SQLITE
    dependencies: [currentHeadcount, previousHeadcount]
    sql: |
      SELECT c.countryCode AS countryCode,
        CAST((c.headcount - p.headcount) AS FLOAT)
             / p.headcount AS growth
      FROM currentHeadcount c
      INNER JOIN previousHeadcount p
        ON c.countryCode = p.countryCode
      ORDER BY growth DESC

target: headcountGrowth
\end{lstlisting}
\caption{The \texttt{HeadcountGrowth} \qdag{}. \code{previousHeadcount} depends on
\code{currentHeadcount} and restricts its query to the countries the first
returned; the \textsc{sqlite} target joins both and computes the growth ratio
(Section~\ref{sec:sql}).}
\label{fig:headcountyaml}
\end{figure}

\subsection{Node types}
\label{sec:nodes}

\qdag{} provides different types of nodes, summarized in Table~\ref{tab:nodes}. They fall
into three groups: \emph{query} nodes that reach external engines (Pinot, RestLi),
\emph{compute} nodes that process JSON in process (JQ, Transform, Join, SQLite,
Literal), and \emph{control} nodes (Conditional, Call). The set is deliberately
open: a node type is an implementation of a small interface
(Section~\ref{sec:engine}), and new engines can be added without touching the
evaluator.

\begin{table}[t]
\centering
\footnotesize
\setlength{\tabcolsep}{4pt}
\caption{The \qdag{} node types, in three groups: query, compute, and
control. Each consumes and produces one JSON value.}
\label{tab:nodes}
\begin{tabular}{@{}l l p{5.0cm}@{}}
\toprule
\textbf{Node} & \textbf{Group} & \textbf{Role} \\
\midrule
\code{PINOT}       & query   & Execute a (templated) OLAP SQL query against a Pinot service and return the result table. \\
\code{RESTLI}      & query   & Issue a downstream service (RESTful) call; project fields. \\
\code{SQLITE}      & compute & Load dependency tables into an in-memory SQLite database and run full SQL (joins, set ops, sub-selects). \\
\code{JOIN}        & compute & Index-based join/merge of two or more table results with inner/outer semantics. \\
\code{JQ}          & compute & Evaluate a jq expression (or YAML macro template) over JSON; the general transform. \\
\code{TRANSFORM}   & compute & Row-wise filter + reshape of a table result into a table or collection response. \\
\code{LITERAL}     & compute & A constant JSON value (e.g.\ a dimension-to-column lookup table). \\
\code{CONDITIONAL} & control & Evaluate a jq predicate and resolve exactly one of two branch nodes. \\
\code{CALL}        & control & Invoke another \qdag{} with a computed input; the unit of reuse. \\
\bottomrule
\end{tabular}
\end{table}

\subsection{Composition and reuse}
\label{sec:reuse}

A \code{CALL} node invokes another \qdag{} by name, passing it an input computed
by a jq expression over local dependencies, exactly as a subroutine call passes
arguments. This is the mechanism by which ``growth rate'' becomes a building block
of many higher-level methodologies, and the lever for the alignment problem of
Section~\ref{sec:motivation}: if two products both \code{CALL} the same canonical
\qdag{}, their definitions cannot silently diverge. Graphs are resolved from a
registry at execution time (Section~\ref{sec:modes}), so a fix to a shared
definition propagates to its callers.

Recursion -- a \qdag{} that calls itself directly or transitively -- is
\emph{representable} (conditionals can terminate it) but is treated as a hazard:
the call chain is tracked, and the design intent is to reject cycles in the call
graph rather than rely on authors to bound them. This is a conservative choice; it
trades expressive power we have not needed for a strong static guarantee of
termination.

\section{The Transformation Sublanguage}
\label{sec:jq}

JSON transformation is the connective tissue of a methodology -- extracting a
field, reshaping rows, computing a ratio, grouping. \qdag{} uses
jq~\cite{jq} as its transformation language (via the JVM \texttt{jackson-jq}
implementation), so a \code{JQ} node is, in the simple case, just a jq expression
with dependency results bound as variables (\code{\$currentHeadcount}, and so on).

jq is expressive but terse, and complex transforms can become difficult to maintain. We therefore
layer a small, optional \emph{macro} surface on top of jq, embedded directly in
YAML, that compiles down to plain jq. The macros (Table~\ref{tab:macros}) trade
jq's density for YAML's structure on the transforms that need it. We are explicit
that this is an ergonomic device, not a new language: it is the well-worn pattern
of a readable structured surface that source-compiles to a terse expression
language, and each macro has a one-line jq expansion.

\begin{table}[t]
\centering
\small
\setlength{\tabcolsep}{6pt}
\renewcommand{\arraystretch}{1.25}
\caption{The optional YAML macro layer over jq: each macro is a source-to-source
expansion, and its leaf values are themselves jq expressions.}
\label{tab:macros}
\begin{tabular}{@{}l p{4.3cm}@{}}
\toprule
\textbf{Macro (YAML)} & \textbf{Expansion (jq)} \\
\midrule
\code{\$do: [e1, e2, ...]}        & \code{(e1) | (e2) | ...} \\
\code{\$let: \{...\} \$do: [...]} & bind variables, then \code{\$do} \\
\code{\$if / \$then / \$else}     & \code{if c then a else b end} \\
\code{\$match / \$cases}          & nested \code{if / elif / else} \\
\code{\$mapEach: e}               & \code{map((e) | select(.!=null))} \\
\code{\$filter: e}                & \code{map(select(e))} \\
\code{\$groupBy / \$groups}       & \code{group\_by(...) | map(...)} \\
\bottomrule
\end{tabular}
\end{table}

\section{Dynamic SQL and the Predicate DSL}
\label{sec:sql}

A query node's \code{sql} field is a template with two placeholder forms, each a
jq expression over the node's resolved dependencies:

\begin{itemize}[leftmargin=1.2em,itemsep=2pt,topsep=2pt]
  \item \textbf{Checked} \texttt{\{\{\,jq\,\}\}} -- a \emph{value}. The result must
  be a scalar or an array of scalars; scalars are quoted and arrays expand to a
  comma list, so a value can never escape its literal position. This is the safe
  default -- e.g.\ the quoted scalar and the \code{IN}-list in
  Figure~\ref{fig:headcountyaml}.
  \item \textbf{Unchecked} \texttt{\{\{!\,jq\,\}\}} -- a \emph{fragment}, for
  injecting a column name or SQL fragment (e.g.\ a dynamic group-by dimension). It
  is spliced verbatim, so the author owns its safety, but is parsed with Apache
  Calcite~\cite{calcite} to reject malformed SQL early.
\end{itemize}

For the common, dangerous case of building a \code{WHERE} clause from optional
request parameters, a small \emph{predicate DSL} lets an author emit a JSON object
of fields and predicates (\code{eq}, \code{lt}, \code{in}, \code{between},
\code{and}/\code{or}/\code{not}) that \qdag{} compiles to a validated fragment,
dropping null sub-predicates so that absent filters simply disappear. A classic
source of injection vulnerabilities~\cite{halfond} becomes a data-construction
problem.

\qdag{} uses Calcite only to parse and validate -- never to optimize, plan, or
push down. It composes opaque query steps; it does not federate them
(Section~\ref{sec:related}).

\section{The Execution Engine}
\label{sec:engine}

The engine's job is to turn a \qdag{} and an input into a result, fast, in the
request path. Its design is a direct, deliberate transplant of the build-system
evaluation model~\cite{feldman,bsalc} into an asynchronous, per-request setting.
A graph is parsed once into immutable, thread-safe node objects (with
pre-compiled jq and SQL templates) and cached, so steady-state execution does no
parsing.

\subsection{Demand-driven, memoized evaluation}
\label{sec:eval}

Evaluation is keyed on the target. An \emph{invocation} object holds the
per-request input and, for each node, a lazily-created future of that node's
result:
\[
\text{node} \;\longmapsto\; \mathit{Lazy}\langle \mathit{Future}\langle
\mathit{Value}\rangle\rangle .
\]
Asking for the target's result forces its lazy future, which forces the futures of
its dependencies, and so on transitively. Two key properties follow from this structure, and they are the whole point:

\begin{itemize}[leftmargin=1.2em,itemsep=2pt,topsep=2pt]
  \item \textbf{Build only what is needed.} A node whose future is never forced
  never runs. Nodes unreachable from the target -- and, crucially, the untaken
  side of a conditional (Section~\ref{sec:cond}) -- cost nothing.
  \item \textbf{Build it at most once.} The lazy future is memoized: a shared
  dependency forced along several paths is computed a single time and its
  result reused. In Figure~\ref{fig:headcount}, \code{currentHeadcount} is forced
  by both \code{previousHeadcount} and the join, but the (potentially expensive)
  Pinot query runs once.
\end{itemize}

Laziness is implemented with a double-checked, lock-guarded holder so that, under
concurrent forcing from independent branches, the underlying supplier runs exactly
once. The result is the classic guarantee of demand-driven memoized DAG
evaluation -- familiar from \texttt{make}, Bazel, and the incremental-computation
literature~\cite{adapton} -- realized here over asynchronous futures and
request-scoped (rather than persistent) memoization.
Algorithm~\ref{alg:eval} states the evaluator precisely.

\begin{algorithm}[t]
\small
\caption{Demand-driven, memoized, parallel evaluation of a \qdag{}.}
\label{alg:eval}
\begin{algorithmic}[1]
\Require graph $G$, target $t$, request input $x$
\State $\textit{inv.cache} \gets \{\}$
  \Comment{node $\mapsto$ \textsc{Lazy}$\langle$\textsc{Future}$\langle$value$\rangle\rangle$}
\State \Return \textsc{Await}(\Call{Force}{$t$})
\Statex
\Function{Force}{$n$}
  \Comment{returns a memoized future; runs $n$ at most once}
  \If{$n \in \textit{inv.cache}$} \Return $\textit{inv.cache}[n]$ \Comment{double-checked, lock-guarded}
  \EndIf
  \State $\textit{inv.cache}[n] \gets \textsc{Lazy}\big(\,\lambda.\ \Call{Eval}{n}\,\big)$
  \State \Return $\textit{inv.cache}[n]$
\EndFunction
\Statex
\Function{Eval}{$n$}
  \If{$n$ is \textsc{Conditional}}
    \State $d \gets \textsc{Await}(\Call{Force}{n.\textit{predDep}})$
    \State $b \gets \textsc{EvalPred}(n.\textit{pred}, d)\ ?\ n.\textit{then} : n.\textit{else}$
    \State \Return $\textsc{Await}(\Call{Force}{b})$
      \Comment{soft dep: untaken branch never forced}
  \Else
    \State $F \gets \{\,\Call{Force}{d} : d \in n.\textit{deps}\,\}$
      \Comment{independent deps in flight together}
    \State $R \gets \textsc{AwaitAll}(F)$
    \State \Return $n.\textsc{Run}(R, x)$
      \Comment{async: Pinot / SQLite / jq / service call / \dots}
  \EndIf
\EndFunction
\end{algorithmic}
\end{algorithm}

Because each node's future is forced at most once, a request performs $O(V+E)$
node forcings for a graph of $V$ nodes and $E$ dependency edges, regardless of how
many paths share a node; the engine's own bookkeeping is therefore linear in the
graph and negligible beside the query and transform work it schedules. Latency is
bounded by the graph's \emph{critical path}, not its size, and conditional pruning
(Section~\ref{sec:cond}) removes entire sub-graphs from both the work and the
critical-path bounds.

\subsection{Parallelism}
\label{sec:parallel}

Because node results are futures, independent branches execute concurrently for
free: resolving a node's dependency set composes their futures and waits on all of
them together, so two Pinot queries with no dependency between them are in flight
at once. The author writes no concurrency code; the parallelism is a consequence
of the declared dependency structure. This is the same observation that motivates
in-process async task frameworks such as ParSeq~\cite{parseq} and Twitter's
Finagle~\cite{finagle}, whose composable futures relate concurrent asynchronous
actions; \qdag{} differs in deriving the task graph from a declarative,
demand-driven, memoizing front-end rather than from combinator code, and in making
shared-node memoization a property of the model rather than the author's
responsibility.

\subsection{Conditional pruning}
\label{sec:cond}

A \code{CONDITIONAL} node evaluates its jq predicate first, and only then forces
the future of the selected branch. The branches are referenced as ``soft''
dependencies that do not participate in the eager dependency resolution, so the
untaken branch -- and the entire sub-graph reachable only through it -- is never
forced and never executes. This is how the talent-pool methodology queries only
the table it needs -- the organization-filtered table or the unfiltered one --
without any imperative branching in application code.

\subsection{In-process relational joins}
\label{sec:sqlite}

Several methodologies need a real join or set operation over intermediate results
-- something jq does awkwardly. A \code{SQLITE} node materializes its named
dependency tables into a fresh in-memory SQLite~\cite{sqlite} database (optionally
declaring primary keys to build indexes for large inputs) and runs arbitrary SQL
over them; table names in the \code{FROM} clause are the dependency names,
extracted by a Calcite visitor. This gives authors the full power of SQL for the
``fetch-then-join'' pattern with zero deployment cost. We note this as a design trade-off rather than a measured result: our workloads have
not yet stressed the join engine hard enough to characterize where a columnar
alternative would win.

\subsection{Caching}
\label{sec:cache}

Query nodes are the expensive nodes, and identical sub-queries recur across
requests and across methodologies. \qdag{} attaches an optional read-aside cache
to \code{PINOT} nodes, keyed on the bound query text, request principal,
pagination parameters, and a cache-configuration name matched (with wildcard
support) against the calling context. Backing storage is a two-tier arrangement:
a local in-process cache (Caffeine~\cite{caffeine}) backed by a distributed blob store. This is an
operational facility rather than a research one, but it matters: because the
cache key is the \emph{fully-bound} query, the memoization of
Section~\ref{sec:eval} (within a request) and the cache (across requests) compose
naturally.

\subsection{Execution modes}
\label{sec:modes}

\begin{figure}[t]
\centering
\includegraphics[width=\columnwidth]{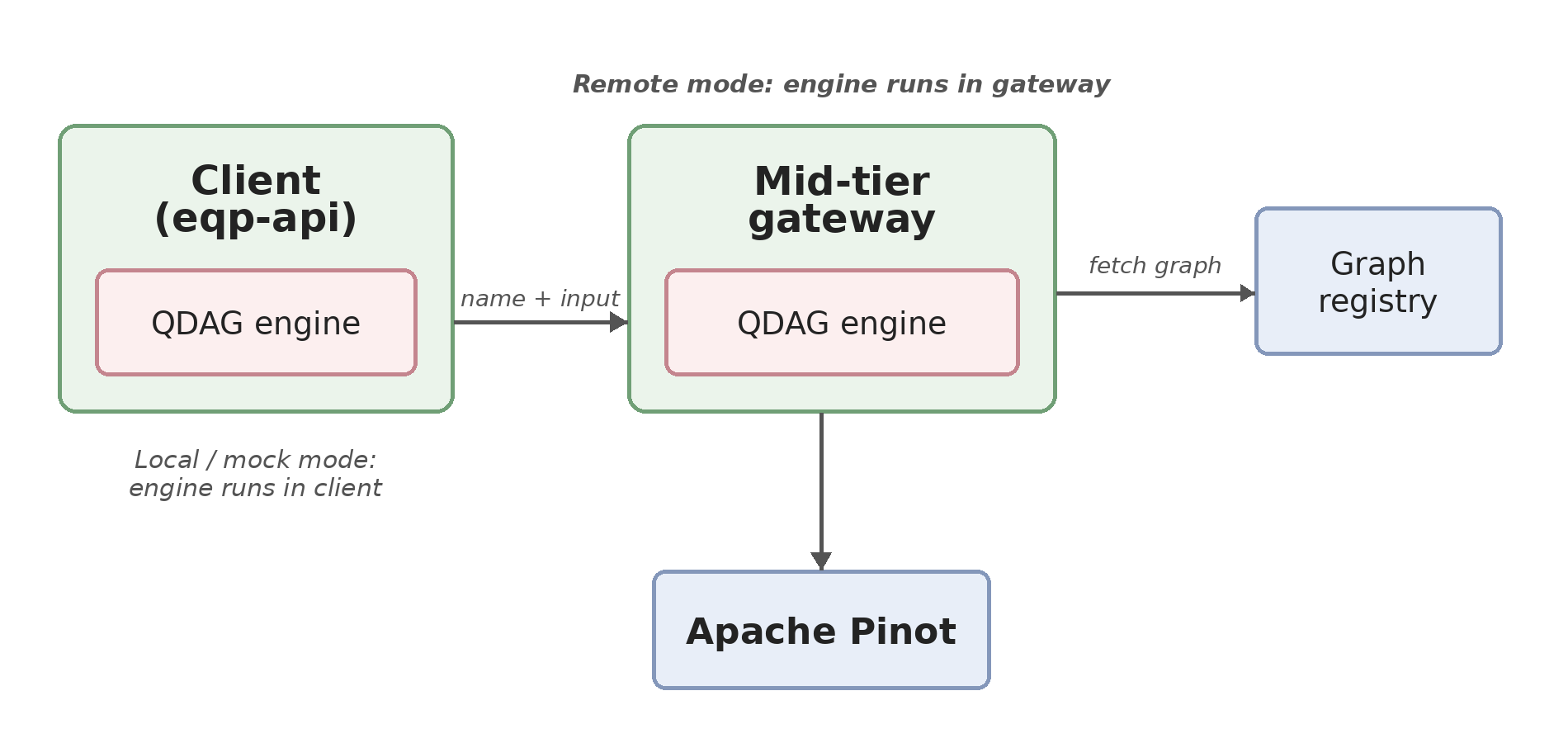}
\caption{The same engine library runs embedded in the client (Local/Mock modes) or
in the mid-tier gateway (Remote/production mode). In production the graph is
fetched from a registry and executed beside the Pinot query path.}
\label{fig:modes}
\end{figure}

The engine is a library, not a service, and the same library runs in two modes
(Figure~\ref{fig:modes}):

\begin{itemize}[leftmargin=1.2em,itemsep=2pt,topsep=2pt]
  \item \textbf{Remote.} The graph is fetched from a registry and executed inside
  the analytics mid-tier gateway, co-located with the Pinot query path; the client
  merely names a \qdag{} and supplies input. This is the production path:
  co-location avoids an extra network hop and reuses the gateway's query
  infrastructure.
  \item \textbf{Local.} The same engine runs \emph{embedded in the client},
  reading graphs from the local classpath, with query nodes reaching data through
  the gateway -- for interactive development. A \emph{mock} variant goes further,
  returning author-supplied tables instead of hitting any data store, so a
  methodology becomes an ordinary unit test: assert that this graph, on this
  mocked data, yields this result.
\end{itemize}

That a methodology can be developed and unit-tested with mock data, before its
Pinot tables even exist, is one of the most-valued properties in practice
(Section~\ref{sec:experience}) -- a direct dividend of making the methodology
data rather than code.

\section{What the Declarative Structure Buys}
\label{sec:analyzable}

Making a methodology data, not code, lets the engine reason about it. Three
capabilities follow, of increasing ambition.

\paragraph{Static validation.}
Before any execution, a graph is checked: every referenced dependency exists, the
target exists, jq and SQL templates parse, and -- by depth-first search -- the
dependency graph is acyclic, with any cycle reported as a concrete path. Errors
that would be runtime exceptions in hand-written glue become load-time rejections.

\paragraph{Testability and observability.}
Because every node is a named, typed step the engine mediates, per-node
instrumentation -- latency, errors, cache hits, attributed to the calling
methodology -- is uniform and free to the author, rather than something each team
adds by hand.


\section{Deployment Experience}
\label{sec:experience}

\qdag{} is implemented as a Java library and is deployed in production in LinkedIn's analytics mid-tier -- across a fleet of more than 500 hosts, with over 600 \qdag{}s defined across more than 100 use cases. We report two kinds of evidence: a controlled pre-rollout
benchmark that isolates the engine's cost, and production telemetry over the
period since.

\begin{figure}[t]\centering
\includegraphics[width=\columnwidth]{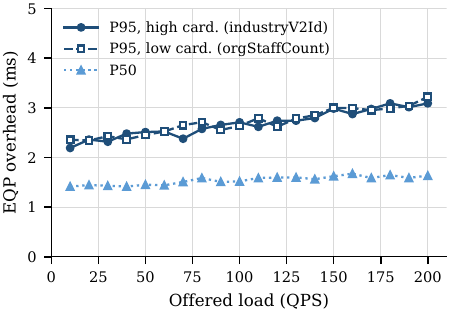}
\caption{Pre-rollout: EQP overhead stays near $2$\,ms across the load range and is
indistinguishable for high- and low-cardinality grouping dimensions.}
\label{fig:perf-overhead}
\end{figure}

\begin{figure}[t]\centering
\includegraphics[width=\columnwidth]{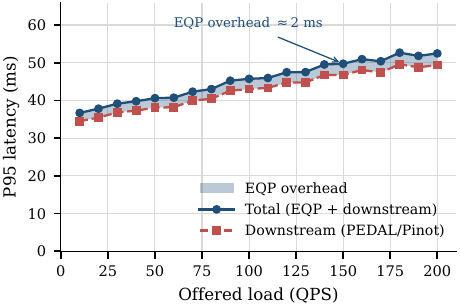}
\caption{Pre-rollout: EQP overhead is a thin band atop downstream PEDAL/Pinot
latency, which dominates end-to-end P95 across the load sweep.}
\label{fig:perf-decomp}
\end{figure}

\paragraph{Pre-rollout validation.}
Before the production rollout we benchmarked the engine on a dedicated perf host
(6 cores, 12\,GiB) against a Pinot cluster (1 broker, 18 servers) mirroring a
production use case. In this controlled, single-use-case setting the engine's
isolated processing overhead -- total time minus the time spent in the parallel
downstream calls -- held near $2$\,ms across the production load range and was
indistinguishable for high- and low-cardinality grouping dimensions
(Figure~\ref{fig:perf-overhead}). Across the load sweep this overhead was a thin
band atop downstream Pinot latency, which dominated end-to-end P95
(Figure~\ref{fig:perf-decomp}). A capacity sweep showed a single host sustaining
$\approx$400\,QPS before throughput plateaus; the limit is JVM thread-pool
saturation while threads wait on downstream calls, not engine CPU, and it
comfortably exceeds the $\approx$300\,QPS-per-fabric requirement.

Because the ceiling reflects blocking threads awaiting downstream responses rather
than engine CPU, a non-blocking or virtual-thread execution model would likely
raise it; we have not measured this, and note it for readers deploying a similar
layer.

\begin{figure}[t]\centering
\includegraphics[width=\columnwidth]{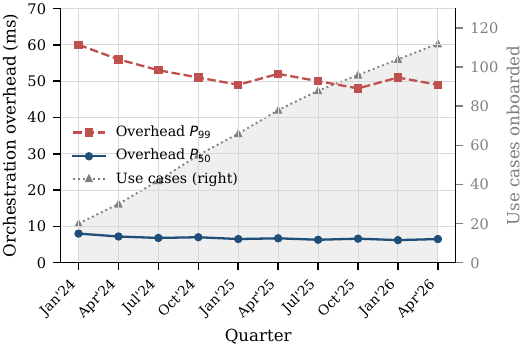}
\caption{\qdag{} orchestration overhead in production. $P_{99}$ starts near
$60$\,ms in early 2024 and settles to $\approx$50\,ms; $P_{50}$ holds in the
$5$--$10$\,ms band -- both stable as adoption grows past 100 use cases (right
axis).}
\label{fig:overhead-trend}
\end{figure}

\paragraph{Production overhead at scale.}
We define \qdag{}'s \emph{orchestration overhead} as the latency it adds on top of
the underlying data-store and downstream-service work: we exclude time spent inside
external Pinot queries and downstream calls and attribute the remainder to graph
resolution, scheduling, memoization, and inter-node transforms. Computed from
production gateway traces across all \qdag{} invocations, this overhead sits in the
$5$--$10$\,ms range at $P_{50}$ and near $50$\,ms at $P_{99}$ -- higher than the
controlled single-use-case microbenchmark above, as expected, since production
percentiles aggregate over far larger and more varied graphs and their tails. As
Figure~\ref{fig:overhead-trend} shows, $P_{99}$ overhead started near $60$\,ms in
early 2024, settled to roughly $50$\,ms within two quarters, and held there, while
$P_{50}$ stayed in the $5$--$10$\,ms band throughout -- both stable even as adoption
grew from a handful of use cases to more than 100. The per-request cost of the
layer did not grow with the number of methodologies it carries.

\paragraph{Adoption.}
Adopting teams reported roughly $60\%$ faster integration -- from ``SQL in hand''
to a correct, parallel, conditionally branching endpoint -- relative to the
hand-written orchestration a \qdag{} replaces. The latency figures are measured in
production; the integration-time figure is adopter-reported, and we treat it as
directional evidence rather than a controlled experiment.The integration-speed figure remains adopter-reported; a controlled authoring
comparison is future work.

\paragraph{Artifact size: a code-level comparison.}
To give the integration-speed figure a concrete artifact-level counterpart, we
compared implementations in one production service. The pre-existing imperative
content-gesture analytics path spans 30 production Java files and $3{,}452$
physical lines of code; its per-finder query-builder, response-mapper, and factory
layer alone accounts for 16 modules and $1{,}199$ Java lines. Three methodologies
expressed as \qdag{}s -- single-post impression analytics, creator article
analytics, and creator video analytics -- total $162$ lines of YAML ($146$
non-comment), against $140$ lines of Java integration code to invoke them. We
report this as \emph{artifact-size} evidence rather than as code eliminated: the
\qdag{} implementations were added alongside the legacy path rather than replacing
it, so these numbers describe the declarative artifact a team writes to express a
methodology, not a completed migration. The comparison is nevertheless indicative
of the authoring burden the declarative form removes: the orchestration, response
mapping, and per-finder plumbing that the Java path spells out module by module do
not appear in the \qdag{} declaration at all.

\paragraph{Migration case studies: latency before and after.}
Two production endpoints migrated from hand-written orchestration to \qdag{}
provide a like-for-like latency comparison. In a follow-recommendation surface,
the impression-count query previously served by a bespoke reader service showed a
$P_{99}$ latency averaging $29.8$\,ms over a five-day window (min $18.6$\,ms, max
$98.6$\,ms); the same methodology expressed as a \qdag{} averaged $23.2$\,ms at
$P_{99}$ (min $10.5$\,ms, max $59.9$\,ms) -- roughly a $22\%$ reduction in average
$P_{99}$, with the worst observed spike falling by nearly half. In a second
service, a compound-key consolidation endpoint fell from an average of
$\approx$$10$\,ms to $\approx$$6$\,ms across the fleet, a $\approx$$40\%$
reduction, with per-host maxima dropping by a larger factor still.

We attribute these gains to structure rather than to faster query execution: the
underlying OLAP queries are unchanged. What changes is that the engine issues
independent queries in parallel and memoizes shared sub-results, where the
hand-written path issued them sequentially, and that composition moves into the
gateway beside the data store, removing a network round-trip per additional
downstream call. The comparison is observational -- these are production
migrations, not a controlled A/B -- but it is a like-for-like measurement of the
same methodology, on the same data, before and after, and it shows that the
orchestration layer's overhead is more than offset by the scheduling it performs.

\paragraph{Reuse through \code{CALL}.}
Because \code{CALL} makes cross-methodology reuse an explicit edge, the registry
itself records how much sharing occurs. Across production from about $600$ \qdag{} the
nodes $252$ \code{CALL} are contained in $152$ caller graphs, targeting $117$ distinct
definitions. Reuse is concentrated rather than incidental: $41$ callees have more
than one caller, and $124$ of the $152$ caller graphs -- over four-fifths --
invoke one of those shared definitions. Most significant for alignment is reuse
that crosses ownership boundaries: $37$ calls from $20$ \qdag{}s target five
externally owned definitions, and the two canonical counting pre- and
post-processors are each invoked by $17$ \qdag{}s spanning four consuming
services. These are definitions that, under the prior model, each consuming team
would have re-implemented -- and that would then have been free to drift apart.

\paragraph{A representative methodology: unique post impressions.}
A concrete illustration is the content-analytics impression methodology of
Figure~\ref{fig:before}: two differentially-private aggregate queries (a
per-dimension count and a total), a non-positive-row filter, and a ratio.
\qdag{} does not implement differential privacy itself; it composes
differentially-private query endpoints and makes the downstream filtering and
ratio computation reusable. As a \qdag{}, the two queries become two \code{PINOT}
nodes the engine runs in parallel, and the ``drop non-positive rows'' and
percentage steps become \code{JQ} nodes, yielding a registered, reusable graph.
The structural change is concrete: two downstream API round-trips collapse to one
-- composition now happens inside the gateway, beside Pinot -- and post-processing
that was Java in the application moves into the declaration. So is the maintenance
argument: when Pinot adds SQL surface (the team's example was \code{HAVING}), it is
usable immediately from the \code{sql} template, with no change to any per-team
query-builder module.

\paragraph{Patterns across methodologies.}
Beyond impressions, the deployed methodologies stress different node types and
validate the generality of the model: talent-pool reporting needs a
\code{CONDITIONAL} to query one of two tables depending on whether an organization
filter is present; top-skills needs \emph{dependent} query nodes (a top-$N$
selection feeding a second round of queries) with a \code{JQ}/\code{TRANSFORM} node
between them to extract the selected entities; headcount growth
(Figure~\ref{fig:headcount}) needs a shared dependency and an in-process
\code{SQLITE} join. One model and one evaluator cover all of them;
Table~\ref{tab:workload} characterizes four representative methodologies
structurally.

\begin{table}[t]
\centering
\footnotesize
\setlength{\tabcolsep}{3pt}
\renewcommand{\arraystretch}{1.2}
\caption{Four representative deployed methodologies (node counts are
representative; the exact count varies with parameters). Between them they exercise
every node group, both dependency shapes (chain and parallel), conditional pruning,
and in-process joins.}
\label{tab:workload}
\begin{tabular}{@{}p{1.7cm} c p{2.0cm} p{1.0cm} c c c@{}}
\toprule
Methodology & Nodes & Node types & Pinot queries
 & \rotatebox[origin=l]{90}{Parallel}
 & \rotatebox[origin=l]{90}{Cond.}
 & \rotatebox[origin=l]{90}{Join} \\
\midrule
Headcount growth        & $3$       & PINOT$\times2$, SQLITE          & $2$ (chain)    & \no  & \no  & \yes \\
Top skills              & $\ge\!3$  & PINOT$\times{\ge}2$, JQ/TRANS. & $\ge\!2$ (dep.) & \no  & \no  & \no  \\
Talent-pool report      & $3$       & CONDITIONAL, PINOT$\times2$    & $1$ of $2$     & \no  & \yes & \no  \\
Unique post impressions & $4$       & PINOT$\times2$, JQ$\times2$    & $2$            & \yes & \no  & \no  \\
\bottomrule
\end{tabular}
\end{table}

\paragraph{Lessons.}
Three lessons stand out for teams considering a similar layer. First, the
abstraction's biggest practical payoff was not raw performance but
\emph{testability and alignment}: mock execution turned methodologies into ordinary
unit tests, and \code{CALL} plus a shared registry gave teams a concrete mechanism
to reuse a definition rather than re-implement it. Second, once the model was
accepted, friction moved entirely to the \emph{tooling around the artifact} -- a
DAG visualizer, a YAML linter, output-schema validation, and richer per-\qdag{}
metrics broken down by calling service -- which we under-invested in early;
adopters also asked for more first-class downstream dependencies (motivating the
\code{GRPC} node) so that \qdag{} could subsume more of the mid-tier. We
interpret these requests as evidence that the core abstraction was useful: it was
accepted, and the friction had moved to its periphery. Third, keeping specialized concerns behind \emph{seams} -- relational joins delegated to an embedded SQL engine, external systems behind typed query nodes -- kept the core small and made the query layer \emph{extensible}: a query node is an interface, so new backends (e.g.\ a Venice~\cite{venice} key-value store) slot in beside Pinot without touching the model or evaluator

\section{Where \qdag{} Sits: Positioning and Related Work}
\label{sec:related}

\qdag{} builds on established techniques from several areas; its contribution is
their combination for a specific setting -- per-request OLAP composition in the
analytics mid-tier. We relate it to each neighbor and note where it differs.
Table~\ref{tab:compare} summarizes.

\begin{table}[t]
\centering
\scriptsize
\setlength{\tabcolsep}{3.5pt}
\renewcommand{\arraystretch}{1.1}
\caption{\qdag{} against neighboring system classes
(\yes{}~yes, \partialmark{}~partial, \no{}~no). The table locates QDAG's
design point rather than ranking systems..}
\label{tab:compare}
\begin{tabular}{@{}p{2.7cm}ccccccc@{}}
\toprule
& \rotatebox[origin=l]{90}{Declarative artifact}
& \rotatebox[origin=l]{90}{Demand-driven + memoized}
& \rotatebox[origin=l]{90}{Per-request, ms, in-process}
& \rotatebox[origin=l]{90}{Multi-engine (SQL+non-SQL)}
& \rotatebox[origin=l]{90}{Arbitrary JSON transforms}
& \rotatebox[origin=l]{90}{Runtime conditional pruning}
& \rotatebox[origin=l]{90}{Recursive composition} \\
\midrule
Workflow DAG schedulers   & \yes & \partialmark & \no  & \yes & \no  & \yes & \partialmark \\
In-process async          & \no  & \partialmark & \yes & \yes & \partialmark & \yes & \yes \\
Build / incremental       & \yes & \yes & \no  & \partialmark & \no  & \yes & \yes \\
Federated SQL / polystore & \yes & \no  & \no  & \yes & \partialmark & \no  & \partialmark \\
Semantic / metrics layers & \yes & \partialmark & \partialmark & \no  & \no  & \no  & \partialmark \\
GraphQL federation        & \yes & \partialmark & \yes & \partialmark & \no  & \partialmark & \yes \\
\midrule
\textbf{\qdag{}}          & \yes & \yes & \yes & \yes & \yes & \yes & \yes \\
\bottomrule
\end{tabular}
\end{table}

\paragraph{Why not adapt an existing system?}
The natural question is why an existing system could not be adapted rather than
built upon. Two adaptations are worth stating explicitly, because they are the ones
we considered. A build system such as Bazel~\cite{bazel} has exactly the evaluation
model we want, but it is architected for a different regime: its unit of work is a
file target rebuilt on a developer's machine or a CI fleet, its memoization is
persistent across invocations, and its scheduling assumes tasks that run for
seconds to minutes. Transplanting it into a request path that must complete in
milliseconds would mean discarding its process model, its persistent cache, and its
file-oriented target abstraction -- that is, everything but the evaluation strategy,
which is precisely the part we do reuse. A semantic layer such as
LookML~\cite{lookml} or MetricFlow~\cite{metricflow} shares our goal of defining a
business definition once, but its artifact compiles to a single SQL query on a
single engine. A methodology that issues two queries where the second is
parameterized by the first's results, filters rows in JSON between them, branches on
request state, or calls a downstream service has no representation in that model at
all; the extension required is not a feature but a change of kind, from a relational
metric compiler to a procedural, multi-engine evaluator. \qdag{} is therefore an
engineering synthesis: it takes the evaluation strategy from one lineage and the
reuse-and-alignment goal from the other, and supplies the piece neither provides --
a per-request, in-process executor over heterogeneous query and transform steps.

\paragraph{Build systems and incremental computation.}
\qdag{}'s evaluator follows a well-established model: demand-driven evaluation from
a target, memoization of shared sub-results, conditional pruning, and parallel
independent branches. This is the model of \texttt{make}~\cite{feldman} and modern
build systems~\cite{bazel,shake}, unified by ``Build Systems \`a la
Carte''~\cite{bsalc} and formalized by the self-adjusting-computation
line~\cite{adapton,acar}. \qdag{} applies it as a request-scoped, non-incremental
special case, over asynchronous futures in the style of in-process task frameworks
such as ParSeq~\cite{parseq} and Finagle~\cite{finagle}. What differs is the
setting, not the evaluation model: \qdag{} runs this model per request, in process,
at interactive latency, over heterogeneous query and transform steps rather than
file or build targets.

\paragraph{Batch dataflow and workflow systems.}
A long line of systems models computation as a dataflow DAG: batch engines such as
MapReduce~\cite{mapreduce}, Dryad~\cite{dryad}, FlumeJava~\cite{flumejava},
Spark~\cite{spark}, and Naiad~\cite{naiad}, the unifying Beam dataflow
model~\cite{dataflowmodel}, and workflow schedulers such as
Airflow~\cite{airflow}, Dagster~\cite{dagster}, and Argo~\cite{argo}. These target
\emph{throughput-oriented, offline or scheduled} execution over large data sets,
where a stage may run for minutes and the unit of work is a partition. \qdag{}
targets the opposite regime -- a single user request, in process, at interactive
latency, where the unit of work is one OLAP query and the whole graph must finish
in milliseconds -- which is why it is demand-driven and request-scoped rather than
batch-scheduled.

\paragraph{Orchestration, not federation.}
Federated and polystore SQL engines -- Trino~\cite{presto}, Calcite~\cite{calcite},
BigDAWG~\cite{bigdawg}, and the SQL layers of cloud engines such as Spark
SQL~\cite{sparksql} and Snowflake~\cite{snowflake} -- take a single declarative
query and let a global optimizer plan across sources, pushing down predicates and
joins. \qdag{} does not plan: it
executes opaque, author-specified steps, shuttles JSON between them, and uses
Calcite only as a front-end parser and validator. The two address different
problems -- \qdag{} is declarative \emph{composition} that may touch several
engines, not cross-source query \emph{planning}.

\paragraph{Semantic and metrics layers.}
\qdag{} shares the goal of the metrics/semantic-layer movement -- dbt's
MetricFlow~\cite{metricflow}, Looker's LookML~\cite{lookml}, Cube~\cite{cube}, and
Malloy~\cite{malloy}:
define a business definition once, reuse it, keep teams aligned. The difference is
the level it operates at. Those systems model metrics relationally and compile to a
single SQL query on one engine; they do not express runtime conditionals over
request state, arbitrary JSON transforms, downstream service calls, or the
sequencing of separate queries with intermediate processing. \qdag{} is procedural
and multi-engine: it sits below the semantic-layer abstraction and admits exactly
that non-relational, multi-step composition.

\paragraph{Per-request mid-tier DAGs.}
GraphQL~\cite{graphql} federation and resolver DAGs~\cite{apollo} also compose work into a response
per request, in a mid-tier, with per-request dataloader memoization -- so a
request-scoped DAG in the mid-tier is itself familiar. \qdag{} differs in the kind
of node it composes: OLAP queries, in-process relational joins, a JSON transform
language, and dynamic SQL, rather than service-field resolvers.

\paragraph{Summary.}
Each property of \qdag{} has a clear precedent: a declarative artifact,
demand-driven memoized evaluation, conditional pruning, parallelism, multi-engine
execution, JSON transforms, and recursive composition. The comparison is intended
to identify the combination of requirements \qdag{} targets, not to suggest that
neighboring systems are incomplete for their own design goals. What is specific
to \qdag{} is the combination -- all of these together, running per request and
in process at interactive latency, specialized for OLAP methodologies. As
Table~\ref{tab:compare} shows, each property is covered by some prior class,
while none covers the union for this setting.


\section{Conclusion and Future Work}
\label{sec:conclusion}
\label{sec:limitations}

Analytics methodologies are small dataflow programs, and writing them as imperative
glue is what makes them expensive to ship, easy to misalign, and hard to analyze.
\qdag{} makes the methodology a declarative artifact -- a graph of typed,
composable steps over heterogeneous engines -- and hands orchestration,
parallelism, and reuse to a shared engine that evaluates the graph demand-driven,
memoized, and pruned, in the request path. Its evaluation model draws on build
systems and its motivation on semantic layers; what is specific to \qdag{} is the
synthesis and the place it runs: a procedural, multi-engine,
transformation-inclusive declarative layer for OLAP methodologies, sitting below
the semantic abstraction and executing per request in the mid-tier.

The experience to date is encouraging. \qdag{} is in production in LinkedIn's
analytics mid-tier across a fleet of more than 500 hosts and adopted by over 100
use cases, with orchestration overhead measured at roughly $10$\,ms at $P_{50}$ and
under $50$\,ms at $P_{99}$, and adopters reporting materially faster integration of
new methodologies. Just as telling as the numbers is where the friction went: once
the model was accepted, the open requests were for tooling around the artifact, not
changes to the abstraction -- the healthy sign that the core idea held.

Several directions follow naturally from making the methodology data rather than
code. The most distinctive is \emph{reasoning about cost before running}. Because a
methodology is a graph, its structure is a cost expression: modelling each node
$i$'s service time as a random variable $T_i$, the target's latency composes
structurally -- a dependency chain adds ($T = T_a + T_b$), parallel branches take
the slowest ($T = \max_i T_i$), and a conditional taking branch $b$ with probability
$p_b$ gives the mixture $T = \sum_b p_b T_b$. Composed over the graph, these would
yield a methodology's latency distribution and expected per-table query load --
inputs to capacity planning and per-methodology SLAs that are unavailable when the
methodology is opaque code. \emph{We have not implemented this.} Two obstacles are
substantive: percentiles do not compose linearly, so the distribution rather than
the percentile must be propagated (the $P_{99}$ of $\max(T_a,T_b)$ generally exceeds
$\max(P_{99}(T_a), P_{99}(T_b))$); and branch probabilities $p_b$, together with the
load multipliers induced when one \qdag{} \code{CALL}s another, are
workload-dependent and must be estimated from production traces. A restricted
version -- propagating distributions over chains and parallel branches for a handful
of production graphs and validating the predicted $P_{99}$ against measurements --
would be the natural first step, and is our main line of ongoing work. The remaining directions track the system's current
boundaries. The in-process join engine is row-oriented, so a columnar embedded
engine~\cite{duckdb} is a better fit for wide aggregations; the surface syntax is
YAML, whose footguns motivate the DAG visualizer, linter, and output-schema
validation adopters have asked for; and broader first-class coverage of downstream
dependencies (the \code{RESTLI} node) would let \qdag{} subsume more of the
mid-tier. The model itself stays deliberately restricted -- an acyclic graph of
single-input/single-output JSON nodes -- which buys strong static guarantees
(termination, validation) at a cost in expressiveness we have not yet needed; that
boundary is a design bet we expect to revisit as workloads demand.

More broadly, \qdag{} is a bet that the right place to fight definitional drift is
\emph{below} the metric, at the procedural composition of heterogeneous steps, and
that an evaluation model proven in build systems transfers cleanly to the
per-request analytics path. The production experience suggests the bet is sound;
the work ahead is to deepen the measurement of what \qdag{} costs and saves, and to
realize the analyzability the declarative form makes possible.


\end{document}